\documentclass[a4paper,fleqn]{cas-dc}
\usepackage{graphicx}
\usepackage{subfigure}
\usepackage{tabularx}
\usepackage[numbers,sort&compress]{natbib}
\usepackage{braket}
\usepackage{array}
\usepackage{caption}
\usepackage{bm} 
\usepackage{listings}
\usepackage{enumitem}
\begin{document}
\let\WriteBookmarks\relax
\def\floatpagepagefraction{1}
\def\textpagefraction{.001}
\shorttitle{}
\captionsetup[figure]{name={Fig.},labelsep=period}

\title [mode = title]{KPROJ: A Program for Unfolding Electronic and Phononic Bands}                      

\author[1]{Jinxin Chen}[type=editor,
                        auid=,bioid=,
                        prefix=,
                        role=,
                        orcid=]

\author[2]{M. Weinert}[type=editor,
                        auid=,bioid=,
                        prefix=,
                        role=,
                        orcid=0000-0002-2263-2960]

\author[1]{Mingxing Chen}[type=editor,
                        auid=,bioid=,
                        prefix=,
                        role=,
                        orcid=0000-0002-5779-3369]

\address[1]{School of Physics and Electronics, Hunan Normal University, Key Laboratory for Matter Microstructure and Function of Hunan Province, Key Laboratory of Low-Dimensional Quantum Structures and
Quantum Control of Ministry of Education, Changsha 410081, China}

\address[2]{Department of Physics, University of Wisconsin, Milwaukee, Wisconsin 53211, USA}

\ead{mxchen@hunnu.edu.cn}

\cormark[1]

\begin{abstract}
We introduce a program named KPROJ that unfolds the electronic and phononic band structure of materials modeled by supercells. The program is based on the $\textit{k}$-projection method, which projects the wavefunction of the supercell onto the ${\textit{k}}$-points in the Brillouin zone of the artificial primitive cell. It allows for obtaining an effective "local" band structure by performing partial integration over the $\textit{k}$-projected wavefunctions, e.g., the unfolded band structure with layer-projection for interfaces and the weighted band structure in the vacuum for slabs. The layer $\textit{k}$-projection is accelerated by a scheme that combines the Fast Fourier Transform (FFT) and the inverse FFT algorithms. It is now interfaced with several first-principles codes based on plane waves such as VASP, Quantum Espresso, and ABINIT. In addition, it also has interfaces with ABACUS, a first-principles simulation package based on numerical atomic basis sets, and PHONOPY, a program for phonon calculations. 

\textbf{Program summary} \\
Program Title: KPROJ \\
Developer’s repository link: https://github.com/mxchen-2020/kproj \\
Licensing provisions: GPLv3.0 \\
Programming language: Fortran 90 \\
\textit{Nature of problem:} \\
Supercells are widely used to model doped systems and interfaces within the framework of first-principles methods. However, the use of supercells causes band folding, which is unfavorable for understanding the effects of doping and interfacing on the band structure of materials. Moreover, the folding also brings difficulties in explaining the results of angle-resolved photoemission spectroscopy experiments. 
\\
\textit{Solution method:}\\
This program is designed to calculate the unfolded band structure for systems modeled by supercells. The unfolding is performed by projecting the wave functions of the supercell onto the $\textit{k}$-points in the BZ of the primitive cell. The projector operator is built by the translation operator and its irreducible representation. The layer $\textit{k}$-projected band structure is obtained by integrating the projected wave function in a selected spatial window, for which the FFT and inverse FFT algorithms are used to accelerate the calculation.

\end{abstract}




\maketitle

\section{Introduction}
First-principles methods based on density functional theory (DFT) have been widely used in the calculation of electronic structures of solid states. Within this framework, supercells are widely used to model doping and interface\cite{intro1,intro2,intro3,intro4,intro5,intro6,intro7,intro8,cdw1,cdw2,cdw3}. Unfortunately, the use of supercells leads to band folding, which hides the nature of the band structure. For example, an indirect band gap may appear as a direct band gap when the valence band maximum (VBM) and the conduction band minimum (CBM) located at different $k$-points are folded to the same $k$-point. As the size of the supercell becomes large, the calculated band structure may look completely different from that of the primitive cell. Such an effect is unfavorable for understanding the effects of chemical doping and interfacing on the band structure of materials. Moreover, the folded bands can hinder the comparison with the results of angle-resolved photoemission spectroscopy (ARPES)\cite{arpes_trouble1,arpes_trouble12}. 

Over the past decades, many efforts have been devoted to developing band unfolding techniques. In the early 1980s, Weinert \textit{et al.}\cite{weinert} obtained the unfolded band structure of Cu$_3$Au by projecting the wave functions onto the $\textit{k}$-points in the Brillouin zone (BZ) of the face-centered cubic primitive cell of Cu, thereby elucidating the effect of alloying on the band structure of Cu. Popescu and Zunger\cite{zunger1,zunger2} proposed a similar method within the framework of plane waves for obtaining an effective band structure for random alloys modeled by large supercells. Similar approaches within the tight binding approximation based on atomic orbitals\cite{lcao1,lcao2,lcao3} and Wannier functions\cite{other_unfold1} were also proposed. So far, several programs have been developed for the band unfolding based on the above-proposed methods\cite{other_unfold2,other_unfold3,other_unfold4,other_unfold5,code1,code2,code3,code4,code5}.

In this communication, we introduce a band unfolding program named KPROJ based on the $\textit{k}$-projection method\cite{k-project}, that uses the projector built by the translation operator and its irreducible representation. One prominent feature of the program is that it allows for calculating the weights of projected wave functions in an artificially defined spatial window. This functionality is ideally suited for interfaces and systems modeled by slabs, for which a layer projection may be needed to extract an effective unfolded band structure for the component systems. An algorithm combining the FFT and back FFT is used to accelerate the layer projection. It is now interfaced with several \textit{ab initio} packages such as VASP\cite{vasp1,vasp2,vasp3}, Quantum Espresso\cite{qe}, ABINIT\cite{abinit}, and ABACUS\cite{abacus}. In addition, it also has interfaces to Phonopy\cite{phonopy}.


\section{The $\textit{k}$-projection method}
\begin{figure}
	\center 
		\includegraphics[width=1\linewidth]{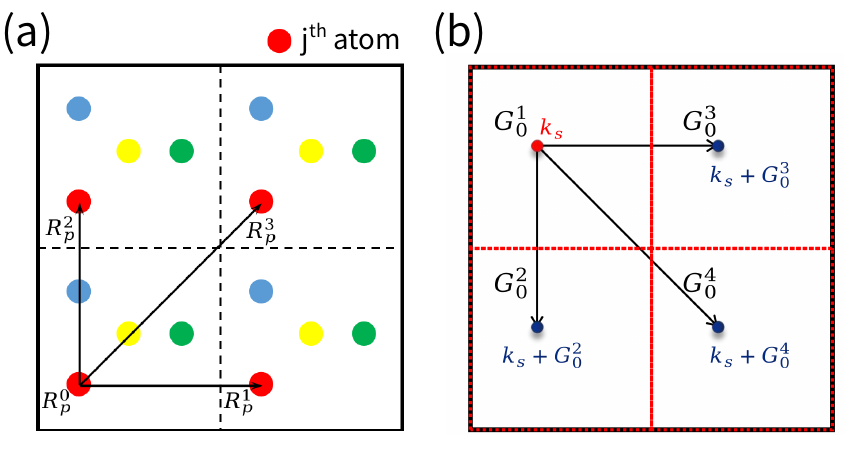}
	\caption{Translation vectors in the real and reciprocal spaces. (a) R$^M_p$ in real space for a $2\times2$ square lattice supercell. (b) Corresponding $G^s_0$ in the reciprocal space.}
    \label{FIG1}
\end{figure}

The details of the $\textit{k}$-projection method have been given in Ref.\  \cite{k-project}. Here we give a brief description of the formalism. Assuming the lattice vectors of a supercell (\textbf{$A_i$}) and the corresponding primitive cell (\textbf{$a_i$}) are related by
\begin{flalign}
&\
\begin{pmatrix}
A_1 \\
A_2 \\
A_3
\end{pmatrix}
=
\mathbf{M}\begin{pmatrix}
a_1 \\
a_2 \\
a_3
\end{pmatrix}
=
\begin{pmatrix}
M_{11} & M_{12} & M_{13} \\
M_{21} & M_{22} & M_{23} \\
M_{31} & M_{32} & M_{33}
\end{pmatrix}
\begin{pmatrix}
a_1 \\
a_2 \\
a_3
\end{pmatrix},
\label{eq1}
\end{flalign} 
where \textbf{M} is the transformation matrix between the lattice vectors of the supercell and the primitive cell. Here, det\textbf{M} = $N_c$, where $N_c$ is the number of unit cells in the supercell. Figure \ref{FIG1}(a) shows $R^M_p$ for a 2 $\times$ 2 supercell of a two-dimensional (2D) square lattice. 

One can also use the matrix \textbf{M} to build the relationship between their reciprocal lattice vectors. In the reciprocal space, multiple $k$-points ($\textbf{k}_p$) in the first BZ of the primitive cell fold back to a single $k$-point ($\textbf{k}_s$) in the first BZ of the supercell
\begin{equation}
\textbf{k}_p = \textbf{k}_s + \textbf{G}^s_0,
\end{equation}
where $\textbf{G}^s_0$ are the reciprocal lattice vectors of the supercell that map the folding. The number of $\textbf{G}^s_0$ equals $N_c$. Fig.\ref{FIG1}(b) schematically illustrates the folding in a 2 $\times$ 2 supercell of the 2D square lattice. The scheme of the $k$-projection band unfolding technique is to build a projector operator($\hat{P}_\textbf{k}$) based on the translation operators ($\hat{\textbf{\textit{T}}}_t$) and the irreducible representations ($\chi_{\boldsymbol{k}_{\mathrm{p}}}$) labeled by $\textbf{k}$ within the first BZ of the primitive cell and apply it to the wavefunctions of the supercell. Then, one needs to calculate the weights of the projected wave functions
\begin{equation}
\hat{P}_k = \frac{1}{h} \sum_t \chi_{\bm{k}}^*(t) \hat{T}_t
\end{equation}
\begin{equation}
\psi_{k_p} = \hat{P}_{\boldsymbol{k}_{\mathrm{p}}}\psi_{k_s}.
\end{equation}
 For wavefunctions expanded in plane waves, the weights of the projected wavefunctions for one of $k_p$ corresponding to a $\textbf{G}^s_0$ are 
\begin{equation}
W_{\textbf{k}_s} = \langle \psi_{\textbf{k}_p} | \psi_{\textbf{k}_p} \rangle = \sum_{\textbf{G}_s} |C^{\textbf{G}_s}_{\textbf{k}_s}|^2,
\end{equation}
where $C^{\textbf{G}_s}_{\textbf{k}_s}$ is the wave function coefficient corresponding to the plane wave $\textbf{k}_s + \textbf{G}_s$, and $\textbf{G}_s$ satisfies
\begin{align*}
\textbf{G}_s &= \sum_i M_i \mathbf{B}_i = \sum_j \left( \sum_i M_i \left( \mathbf{B}_i \cdot \mathbf{a}_j \right) \right) \mathbf{b}_j
\\ &  = \mathbf{G}_p + \textbf{G}_s^0 .
\end{align*}
In the linear combination of atomic orbital approximation, the weights are of the form 
\begin{equation}
    W_{\textbf{k}_p} = \frac{1}{{N_c}} |\sum_M C_{\textbf{k}_s,i,\alpha}^M e^{i \mathbf{k_s} \cdot \mathbf{R_p^M}}|^2
\end{equation}
where $i\alpha$ index the orbital $\alpha$ of i-th atom, and $C_{k_s,i,\alpha}$ is wave function coefficient corresponding to that atomic orbital.   \par
For phonon calculations, the weights are 
\begin{equation}
   W_{\textbf{q}_p} = \frac{1}{N_c} \sum_{\mathbf{R_p^M}} \sum_{\alpha \beta} C_{a j}^* C_{\beta j'} e^{i \mathbf{q_s} \cdot \mathbf{R_p^M}} \delta_{\alpha \beta} \delta_{j j'}
\end{equation} 
where $C_{a j}$ now represents the coefficient of the vibration for $j$-th atom in $\alpha$ direction.
\par
\begin{figure}
	\center 
		\includegraphics[width=0.7\linewidth]{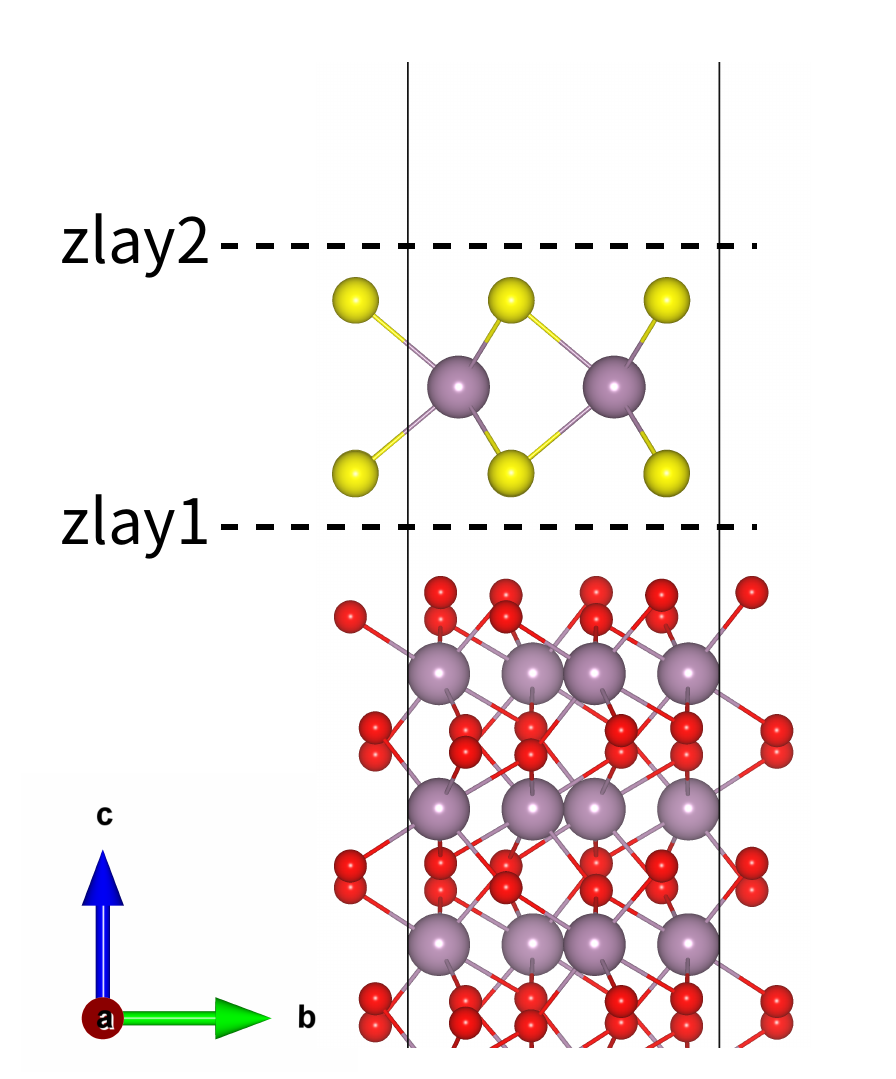}
	\caption{The setting of zlay1 and zlay2 for layer projections is defined by fractional coordinates based on the c-axis.}
    \label{FIG2}
\end{figure}

For interfaces, the two components may be in different supercells. Therefore, one has to project the wave functions of the supercell onto appropriate $k$-points in the first BZ of the interested system. The effective unfolded band structure for the target component can be obtained by a layer $k$-projection, for which an integral over the projected wave function in a specified spatial window is performed, i.e.,
\begin{equation}
A_{\textbf{k}_p} =\int_{zlay1}^{zlay2}\psi_{\textbf{k}_p}^*\psi_{\textbf{k}_p}dr 
\end{equation} 
where zlay1 and zlay2 are the upper and lower bounds of the integral along \textit{z}, see Fig.\ref{FIG2}. 
\par
We use an algorithm that combines FFT and the inverse FFT to accelerate the above calculations.

\section{Description of the KPROJ program}

\subsection{Installation}
The KPROJ includes a Makefile for the Intel Fortran compiler(\textit{\textbf{ifort}}). Below, we outline the detailed installation process, which can also be found in the KPROJ tutorial.\\
(i) First, users should install the FFTW and add its include directory to the Makefile.

\begin{lstlisting}
cd /kproj/src
vim Makeile
Modify the FFTW_INCLUDES parameter :
FFTW_INCLUDES = /your_fftw_directory/include
\end{lstlisting}
(ii) Compile the XML library under ~/kproj/src:
\begin{lstlisting}
make fox
\end{lstlisting}
(iii) Compile the program using \textit{make}. Once the program has been successfully compiled, an executable named \textbf{\textit{kproj}} will be found.
\begin{lstlisting}
make
\end{lstlisting}
\subsection{Framework of the code}
\begin{figure}
	\centering
		\includegraphics[width=1\linewidth]{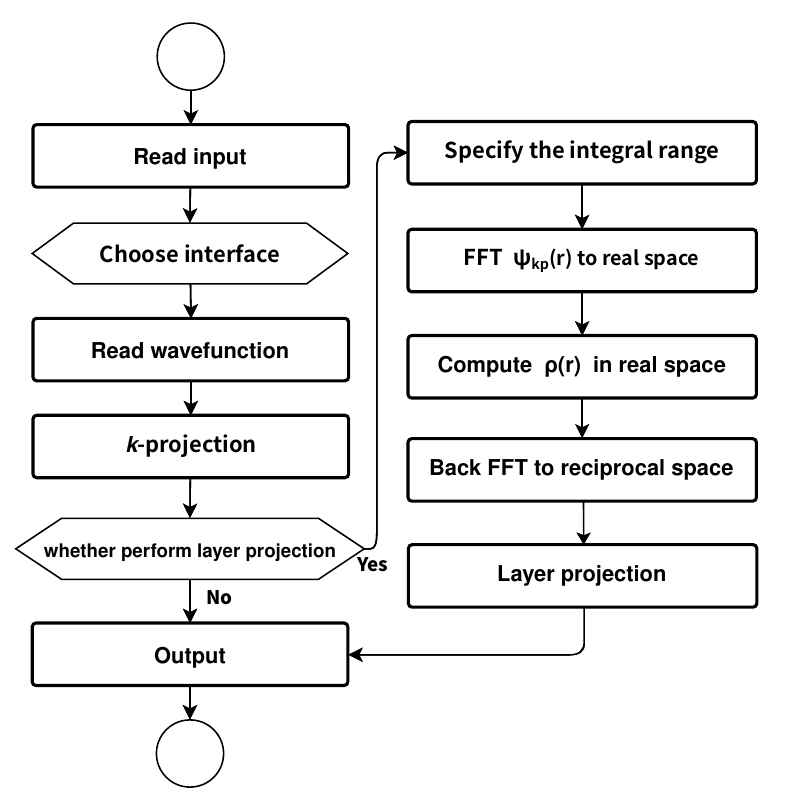}
	\caption{Work flow of KPROJ.}
        \label{FIG3}
\end{figure}

The above formalism is implemented in the KPROJ program, available at Github (\url{https://github.com/mxchen-2020/kproj}). 
The source code is written in Fortran90: (i) main.f controls the workflow of the whole program; (ii) input.f reads the input parameters; (iii) mod$\_$wave.f reads the wave function; (iv) mod$\_$kproj.f performs band unfolding and layer projection calculation; (v) out$\_$bands.f outputs the weights of $\textit{k}$-projected wave functions.
Because the parameters for the wave functions generated by Quantum Espresso are stored in the XML format, one has to install the XML library, which can be done by the terminal command "\textit{make fox}". The executable file of KRPOJ can be obtained after compiling using "\textit{make}". The workflow of KPROJ is shown in FIG.\ref{FIG3}.

\subsection{Inputs and outputs}
The KPROJ program requires a few input files for different interfaces, which are listed in Table \ref{tbl1}.

\begin{table}[h]
\caption{Required input files for different interfaces}
\label{tbl1}
	\centering
	\begin{tabular}{p{3cm}@{}ccc}
     \toprule
     Interface & Input file(s)  & Wavefunction \\
     \toprule
     VASP &  INKPROJ & WAVECAR  \\
      \toprule
     Quantum Espresso & INKPROJ & WFC${\underline{~~}}$i.dat  \\
      \toprule
     ABINIT     &  INKPROJ  & WFCAB \\   
      \toprule
     Phonopy    &  INKPROJ  &  band.yaml  \\
                & pos${\underline{~~}}$map.dat \\
     \toprule
     ABACUS     & INKPROJ  & LOWF${\underline{~~}}$K${\underline{~~}}$i.dat \\
                & pos${\underline{~~}}$map.dat \\
    \toprule
	\end{tabular}
\end{table}

\subsection{INKPROJ}
\begin{figure}
	\centering
		\includegraphics[width=1\linewidth]{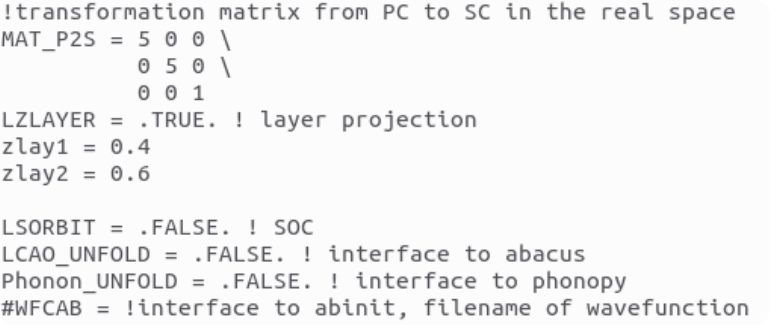}
	\caption{An example of the input file INKPROJ for the KPROJ.}
        \label{FIG4}
\end{figure}

INKPROJ provides the basic input parameters for band unfolding calculations, which can be as simple as the transformation matrix and optional information about the spatial projection; an example with additional parameters is given in Fig.\ \ref{FIG4}. A more detailed description of the input parameters can be found in the folder $doc$.

\subsection{pos$\_$map.dat}
\begin{figure}
	\centering
		\includegraphics[width=1\linewidth]{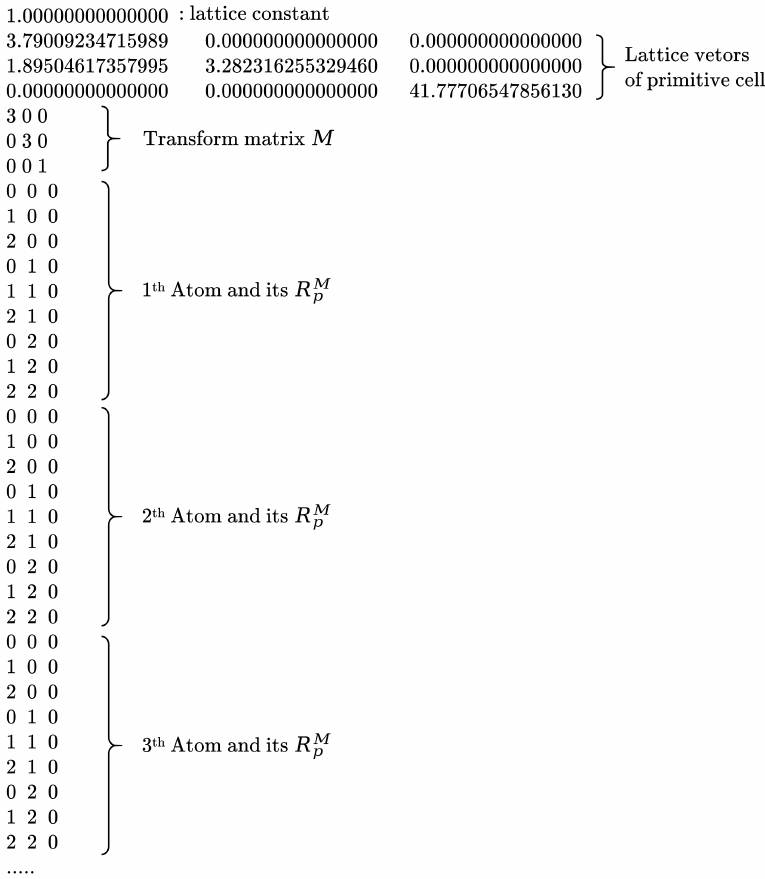}
	\caption{The pos$\_$map.dat for $3\times3$ silicene. The equivalent atoms are placed together and the $R_p^M$ are given.}
      \label{FIG5}
\end{figure}

The file pos$\_$map.dat is only used for the LCAO methods. It stores the lattice vectors $\textbf{a}_i$ of the primitive cell, the transformation matrix $M$, and corresponding translation vectors $R_p^M$ from the primitive cell to the supercell. Figure \ref{FIG5} shows an example for a 3 $\times$ 3 supercell of silicene. We provide a program called "\textit{supercell}" in the folder $utils$ to help users generate supercells and the file pos$\_$map.dat.

\subsection{Output files}
The KPROJ package has two output files, a log file named OUTKPROJ, and  bs$\_$projected.dat, which stores energy eigenvalues and the calculated weights of the $k$-projected wavefunctions. We provide a post-processing program called $futils$, which uses bs$\_$projected.dat to prepare files for plotting in formats such as xmgrace, gnuplot, and opendx.

\subsection{Utilities}
There are a few utilities that can facilitate the use of KPROJ. \\
(i) \textit{supercell} is a program for generating supercells and the file pos$\_$map.dat. Users are only required to provide the transformation matrix $M$ in Eq.~(\ref{eq1}). It is in the directory \textit{Utils} and the executable file can be obtained after \textit{make}.\\
(ii) \textit{kpts\_path} is also in the directory of $Utils$ and used to generate the $k$-points along a high symmetry line defined by two $k$-points, i.e., $k_i$ and $k_f$, which are in the direct coordinate. \\
(iii) \textit{ksplit} is designed to split the $k$-points in a KPOINTS file into a few files named KPOINTS$\_i$, which can be used for heavy calculations. \\
(iv) \textit{mergefile} combines several $bs\_projected.dat$ files named $bs\_projected.dat\_i$ into one file. Note that the ordering should be consistent with the $k$-points in KPOINTS. It is in the \textit{src} directory and is obtained by \textit{make mergefile}. \\
(v) \textit{futils} prepares files for plotting using the obtained $bs\_projected.dat$, which can be obtained by \textit{make futils} in \textit{src} .

\subsection{Typical Workflow} 
In this subsection, we will introduce the steps to run KPROJ. \\
(i) Generate the supercell. \\
 We strongly recommend using \textit{\textbf{supercell}} program to generate supercells. It can automatically generate the pos\_map.dat file, which is an important input file for unfolding the band structures calculated by ABACUS and Phonopy. 
\begin{lstlisting}
./supercell
Transformation matrix (must be integers):
   M(1,1)  M(2,1)  M(3,1)
   M(1,2)  M(2,2)  M(3,2)
   M(1,3)  M(2,3)  M(3,3)
   2 0 0
   0 2 0
   0 0 1
The supercell contains   4  primitive cells
Congratulations: the supercell is 
generated successfully
\end{lstlisting}
(ii) Band calculation (using DFT code). \\
Note that the wave functions should be output and saved during the band calculation. \\
(iii) KPROJ calculation. \\
Users should prepare INKPROJ under the folder which stores the wavefunction. For interfaces/heterostructures, users can set proper values for LZLAYER, zlay1, zlay2 to 
perform layer projection (Currently, this feature is only available for the wavefunction in planewaves, i.e., generated by VASP, Quantum Espresso, and Ab-init). Once everything is prepared, execute the \textbf{\textit{kproj}} to perform \textbf{\textit{k}}-projection.
\begin{lstlisting}
./kproj
\end{lstlisting}
(iv) Prepare files for plotting. \\
Set the correct Fermi energy in bs\_projected.dat. Run \textbf{\textit{futils}} to modify parameters and generate files for different plotting tools. 
\begin{lstlisting}
./futils
-------- k-projected bands ------
OpenDx plot defaults:
emin = -17.000000 (data: -16.742051)
emax =  4.000000  (data:  3.978691)
spacing in energy:
de = 0.050000
minimum weight for any state:
wtmin = 0.000000
energy broadening of state:
dele = 0.050000
ratio of y to x of plot:
y2x =  1.000000
wtcut = 0.000000
if you wish to change, then name=value
to end, "x" on separate line
> emax = 4
emax = 4.000000
> emin = -17
emin = -17.000000
> x
\end{lstlisting}
(v) Plotting \\
Users can use xmgrace, gnuplot, and opendx to plot the unfolded bands.
\subsection{KPROJ for large systems}
\begin{figure}
	\centering
		\includegraphics[width=1\linewidth]{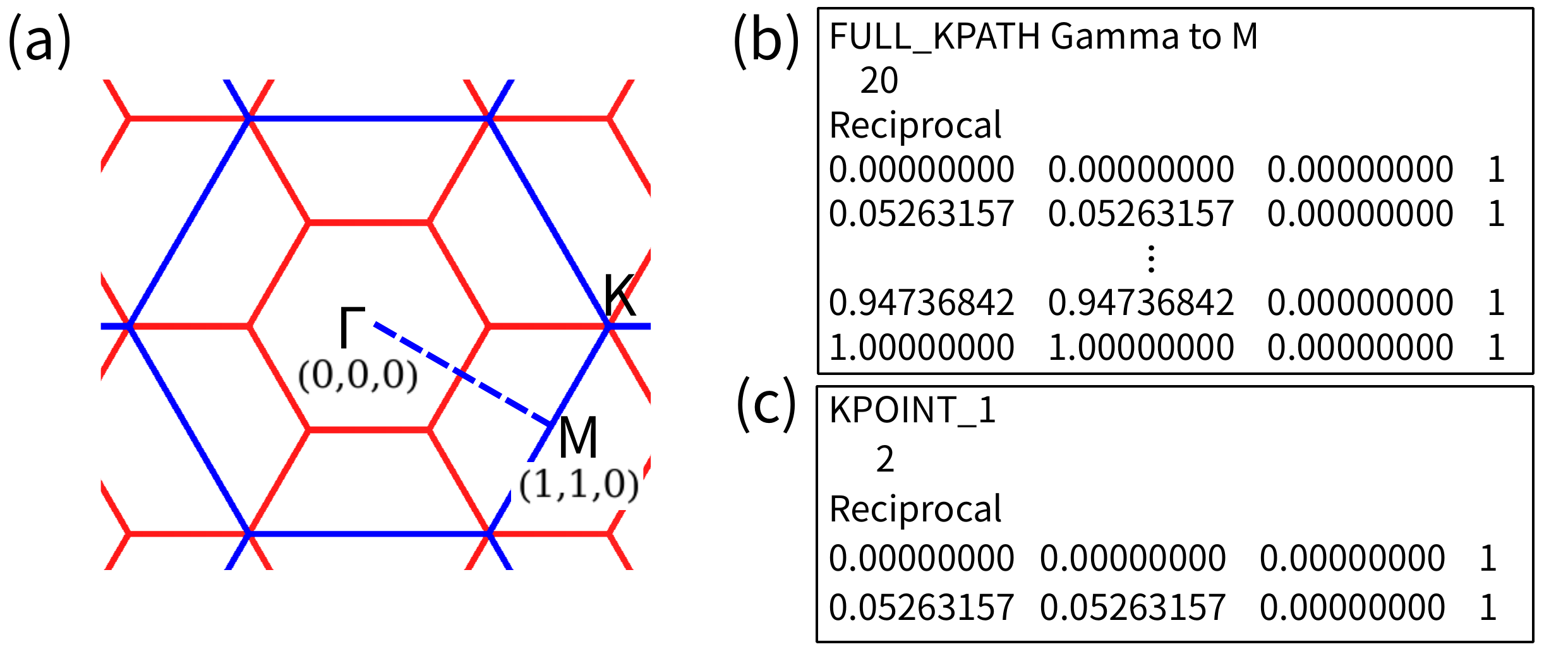}
	\caption{(a) The first BZs of the primitive cell (blue hexagon) and the $2$ × $2$ supercell (red hexagons). (b) The full \textbf{k}-path in IBZKPT format generated by \textit{\textbf{kpts\_path}}. (c) KPOINTS file for distributed computing } 
      \label{FIG6}
\end{figure}

Band calculations with dense $k$-point sampling along multiple directions can require considerable computational memory for large systems($\sim10^3$ atoms). To deal with this issue, we split the full \textbf{k}-path into multiple files, i.e., \textbf{KPOINTS\_i}. Then, we perform individual calculations for each \textbf{KPOINTS\_i}. The detailed steps are as follows: \\
(i) Select the high symmetry path, see Fig.\ \ref{FIG6}(a).  \\
(ii) The full \textbf{k}-path is generated according to the starting and ending \textbf{k} points use \textbf{\textit{kpts\_path}}, shown in Fig.\ \ref{FIG6}(b).
\begin{lstlisting}
./kpts_path
Please type ki(1),ki(2),ki(3) and 
kf(1),kf(2),kf(3)
0.0000000000   0.0000000000   0.0000000000
1.0000000000   1.0000000000   0.0000000000
How many kpoints do you want along the path?
20
\end{lstlisting}
(iii) Using \textit{\textbf{ksplit}} to split the full \textbf{k}-path into several segments shown in Fig.\ref{FIG6}(c)
\begin{lstlisting}
./ksplit
Name of the file containing k-points:
kpts.dat
No. of kpts for each job
2
ls KPONTS_*
KPOINTS_1  KPOINTS_2  KPOINTS_3  KPOINTS_4 
KPOINTS_5  KPOINTS_6  KPOINTS_7  KPOINTS_8
KPOINTS_9  KPOINTS_10
\end{lstlisting}
(iv) Perform the band calculations using KPOINTS\_i and \textbf{\textit{k}}-projection to get the corresponding bs\_projected.dat.  \\
(v) Then, create a new folder, copy (or link) each bs\_projected.dat files here, and rename them to bs\_projected.dat\_i in the order of KPOINTS\_i. Finally, we use \textbf{\textit{mergefile}} to merge all bs\_projected.dat\_i.
\begin{lstlisting}
./mergefile
No. of eigenvalue files
20
Mode 1: merge k-points
Mode 2: merge bands
1
Flip bands around the first kpt?
0 for No and 1 for yes
0
\end{lstlisting}
\subsection{Benchmark}
\begin{figure}
	\centering
		\includegraphics[width=1\linewidth]{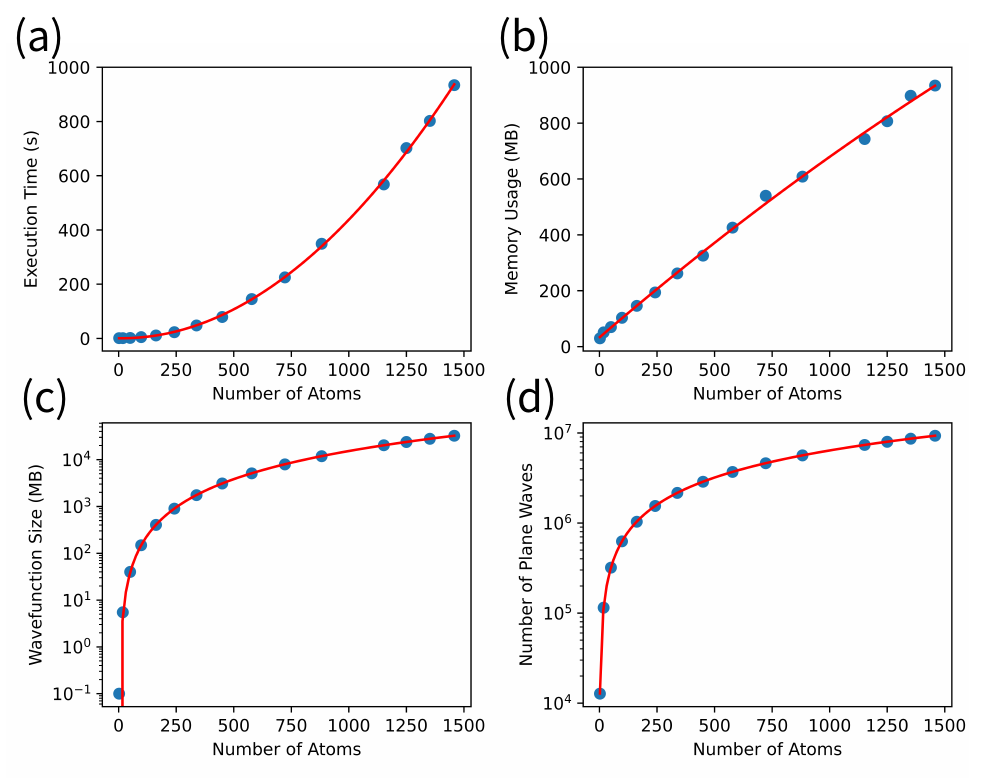}
	\caption{Numbers of atoms vs.\ (a) CPU hours; (b) Memory usage; (c) Wavefunction size; (d) Number of plane waves} 
     \label{FIG7}
\end{figure}

We perform layer \textit{\textbf{k}}-projection for a single \textbf{k} point on a node with 8 cores and 16 threads. An energy cut-off of 500 eV was used for the electronic structure calculation. For instance, for a graphene supercell with about 1000 atoms, the layer \textbf{\textit{k}}-projection can be completed within ten minutes. During the unfolding process, KPROJ only reads the plane waves from one band of the wavefunction. As a result, the memory consumption of layer \textbf{\textit{k}}-projection is remarkably low, requiring only 1 GB, as illustrated in Fig.\ \ref{FIG7}(b). Additionally, we have also compared our code with other band unfolding programs, see Table \ref{tbl2}. 

\begin{table*}[h]
\caption{Comparison with other codes.}
\label{tbl2}
	\centering
	\begin{tabular}{p{4cm}@{}ccc}
     \toprule
     Program & Interface(s)  & Function(s) \\
     \toprule
     upho\cite{upho} &  Phonopy & Phononic band unfolding  \\
      \toprule
     BandUp\cite{BandUp} & VASP,QE,AB-init & Electronic band unfolding \\
      \toprule
     PYATB\cite{PYATB}     &  ABACUS  & Electronic band unfolding \\   
      \toprule
     Pyprocar\cite{pyprocar}    &  VASP  &  Electronic band unfolding  \\
     \toprule
     fold2Bloch\cite{fold2Bloch}     & VASP,wien2k & Electronic band unfolding \\
    \toprule
    VASPKIT\cite{vaspkit}       & VASP & Electronic band unfolding \\
    \toprule
    Phonon Unfolding\cite{Phononunfolding}  & Phonopy & Phononic band unfolding \\
    \toprule
    Quantum Unfolding\cite{quantumunfolding}  & wannier90 & Electronic band unfolding \\
    \toprule
    KPROJ\cite{k-project}  & VASP,QE,AB-INIT,ABACUS,Phonopy & Electronic and Phononic band unfolding\\   & & Layer-projection \\
    \toprule
	\end{tabular}
\end{table*}

\section{Applications}
In the following, we show a few applications of the code in revealing the effects of doping and interfacing on the band structures of materials.

\subsection{Graphene supercells}
\begin{figure}
	\centering
		\includegraphics[width=1\linewidth]{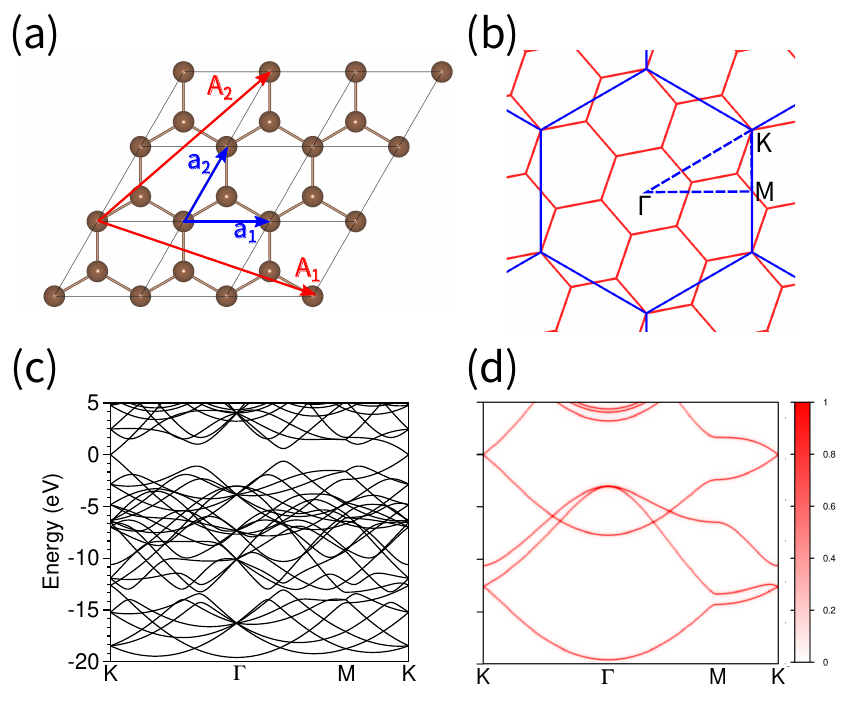}
	\caption{$\textit{k}$-projected (unfolded) band structure for perfect graphene. (a) Geometric structure of graphene in a $\sqrt{7}$ × $\sqrt{7}$ supercell. The lattice vectors of the primitive cell and the supercell are shown in blue and red, respectively. (b) The first BZs of the primitive cell (blue hexagon) and the $\sqrt{7}$ × $\sqrt{7}$ supercell (red hexagons). High symmetry points are labeled. (c, d) The band structures without and with band unfolding along the dashed lines shown in (b).} 
     \label{FIG8}
\end{figure}

We start with perfect graphene in a $\sqrt{7}$ × $\sqrt{7}$ supercell. Figure \ref{FIG8}(a) shows the geometric structure, and Fig.\ \ref{FIG8}(b) the BZs, of the primitive cell and supercell. Figure \ref{FIG8}(c) shows the band structure along the high symmetry lines in the BZ of the primitive cell directly obtained from the supercell DFT calculations. Applying KPROJ to these states yields the unfolded band structure shown in Fig.~\ref{FIG8}(d), which is identical to the one calculated using the primitive cell.

\subsection{Spin-orbit coupling}
\begin{figure}
	\centering
		\includegraphics[width=1\linewidth]{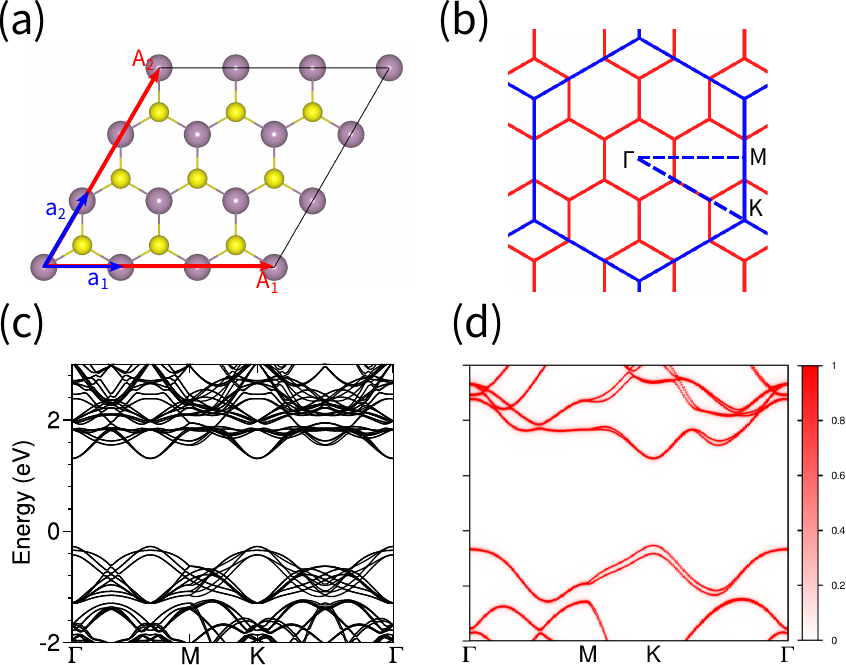}
	\caption{$\textit{k}$-projected (unfolded) band structure for perfect MoS$_2$. (a) Geometric structure of graphene in a 3 × 3 supercell. The lattice vectors of the primitive cell and the supercell are shown in blue and red, respectively. (b) The first BZs of the primitive cell (blue hexagon) and the 3 × 3 supercell (red hexagons). High symmetry points are labeled. (c, d) The band structures without and with band unfolding along the dashed lines shown in (b).} 
     \label{FIG9}
\end{figure}

In transition metal dichalcogenides(TMDCs), spin-orbit coupling (SOC) plays a crucial role, and is essential for valleytronics. In this subsection, we apply KPROJ to MoS$_2$ in a 3 × 3 supercell with SOC calculations. The geometric structure is shown in Fig.~\ref{FIG9}(a), for which the BZs of the supercell and the primitive cell is shown in Fig.~\ref{FIG9}(b).
Figure~\ref{FIG9}(c) shows the calculated (supercell) bands along the high-symmetry lines shown in Fig.\ \ref{FIG9}(b). Because of the band folding, the band splitting at K is hidden. After the \textbf{\textit{k}}-projection calculations, the band splitting induced by spin-orbit coupling (SOC) can be clearly observed at the K point, as shown in Fig.~\ref{FIG9}(d).

\subsection{Defected graphene}
\label{defected_graphene}
\begin{figure}
	\centering
		\includegraphics[width=1\linewidth]{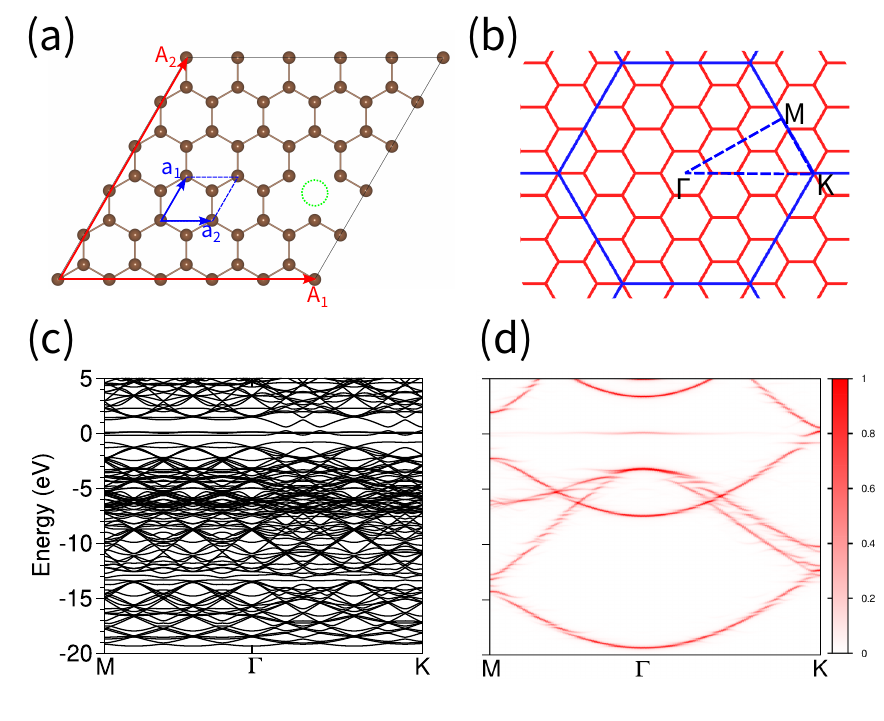}
	\caption{$\textit{k}$-projected band structure for defected graphene. (a) Top view of the structure with one defect in a 5 $\times$ 5 supercell. (b) The first BZs of the primitive cell (blue) and the supercell (red). High symmetry lines are shown in blue dashed lines. (c, d) respectively show the band structures without and with band unfoldings.}
  \label{FIG10}
\end{figure}

Doping and defects play an important role in tuning the electronic properties of materials\cite{doping1,doping2,doping3,doping4,doping5,doping6}. Within the framework of first-principles calculations, such systems are usually modeled by supercells. Figure \ref{FIG10}(a) shows the geometric structure of the defected graphene derived from a 5 $\times$ 5 supercell, for which the BZ is shown in Fig.~\ref{FIG10}(b). Figure \ref{FIG10}(c) shows the band structure for the defected 5 $\times$ 5 graphene. Besides the folding, which makes it difficult to see the graphene bands, there are two new features: dispersionless bands near the Fermi level (as expected for a localized isolated defect), and the apparent disappearance of the Dirac cone. By projecting the wavefunction of the defected graphene onto the $k$-points in the first BZ of the corresponding primitive cell, (i) the bulk graphene states are recovered; (ii) there are two defect states, one is distributed over the whole BZ, and the other lower one is in the gap opened up in the lower part of the Dirac cone, and is mainly derived from those pure graphen states; and (iii) there are additional minigaps in the electronic bands due to defects. 

\subsection{Graphene bilayer on 6H-SiC}
\begin{figure*}
	\centering
		\includegraphics[width=0.95\linewidth]{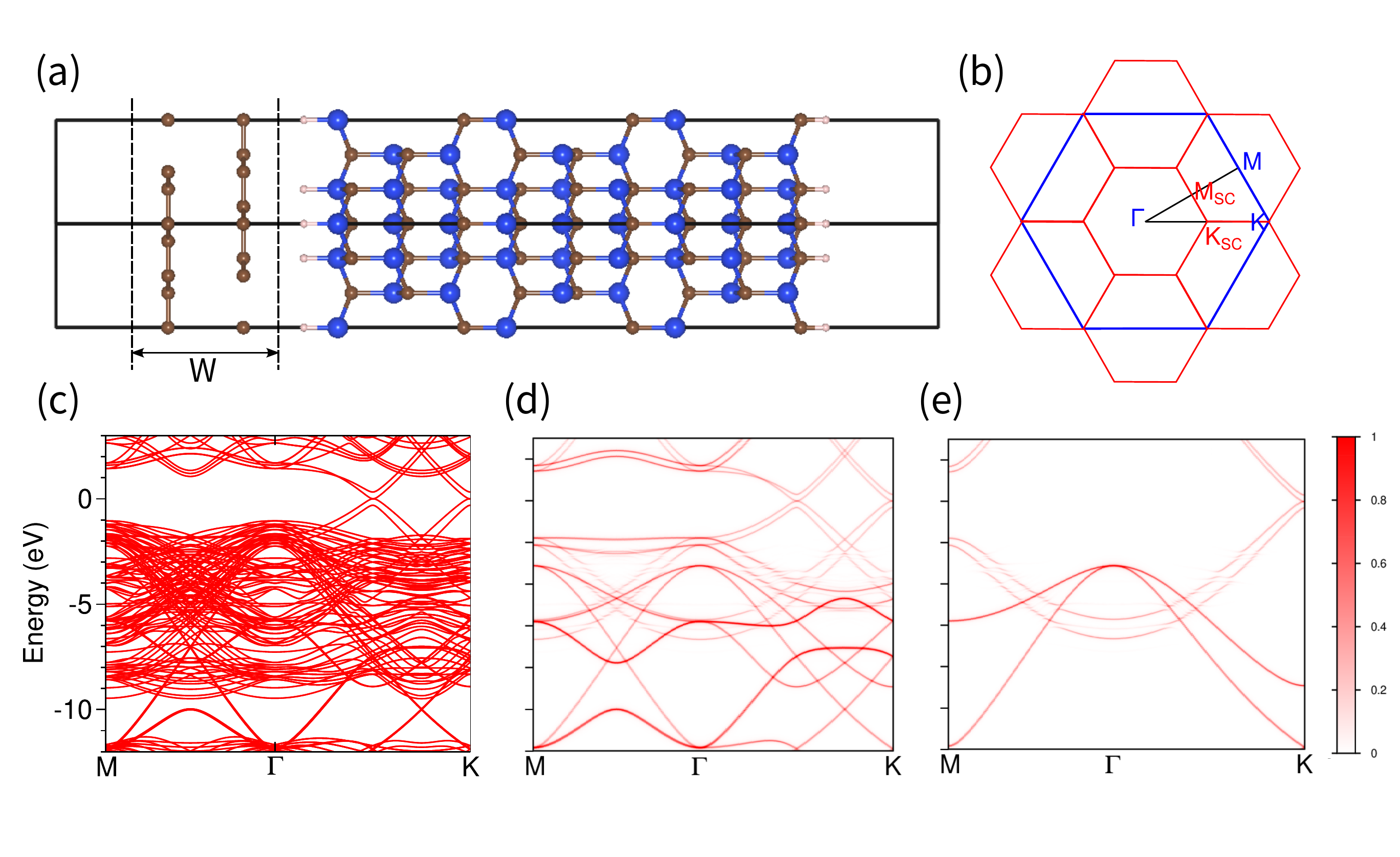}
	\caption{$\textit{k}$-projected band structure for graphene-2L/6H-SiC(0001). (a) Geometric structure of the interface structure. The wavefunctions in the spatial window defined by $W$ are selected for the $\textit{k}$-projection calculations for graphene-2L. (b) The first BZs of graphene-2L/6H-SiC(0001) and the primitive cell of graphene. The high symmetry points and lines used for the calculations are shown. (c) The band structure obtained from a standard VASP calculation. (d) The band structure weighted by contributions of the graphene-2L. (e) The $k$-projected band structure for the graphene-2L.}
  \label{FIG11}
\end{figure*}

Now we show an application of KPROJ in interfaces. We performed calculations for a graphene bilayer (graphene-2L) on 6H-SiC(0001), which has two types of terminations, i.e., Si and C terminations\cite{SiC1,SiC2,SiC3,SiC4,SiC5,SiC6,SiC7,SiC8}. Here, we consider an interface structure consisting of a 2 $\times$ 2 supercell of the graphene-2L on a $\sqrt{3} \times \sqrt{3}$ Si-terminated 6H-SiC(0001), which is saturated by hydrogen atoms. The geometric structure is shown in Fig.\ref{FIG11}(a), for which the BZs are shown in Fig.\ref{FIG11}(b). The calculated bands along the BZ of the graphene primitive cell are shown in Fig.\ref{FIG11}(c). Note that there are two sets of band crossings along the path from $\Gamma$-K. One is located at a $k$-point (named $k_q$) in the middle of $\Gamma$-K and the other emerges at K. The extra band crossing at $k_q$ is due to band folding, which still exists in the band structure weighted by the contribution of graphene shown in Fig.~\ref{FIG11}(d). Then, we perform $k$-projection over the wave functions in the overlayer. The wave functions in the spatial window defined in Fig.~\ref{FIG11}(a) are chosen for the calculations. Then one can obtain an "effective" band structure for the bilayer graphene. The main profile of the unfolded band structure shows great similarity with that of the freestanding systems, except for minigaps in the electronic bands. Therefore, one can deduce that the Si-terminated 6H-SiC(0001) with H-saturation has a minor effect on the band structure of the overlayer. 

\subsection{Bulk projection of surfaces}
\begin{figure*}
	\centering
		\includegraphics[width=1\linewidth]{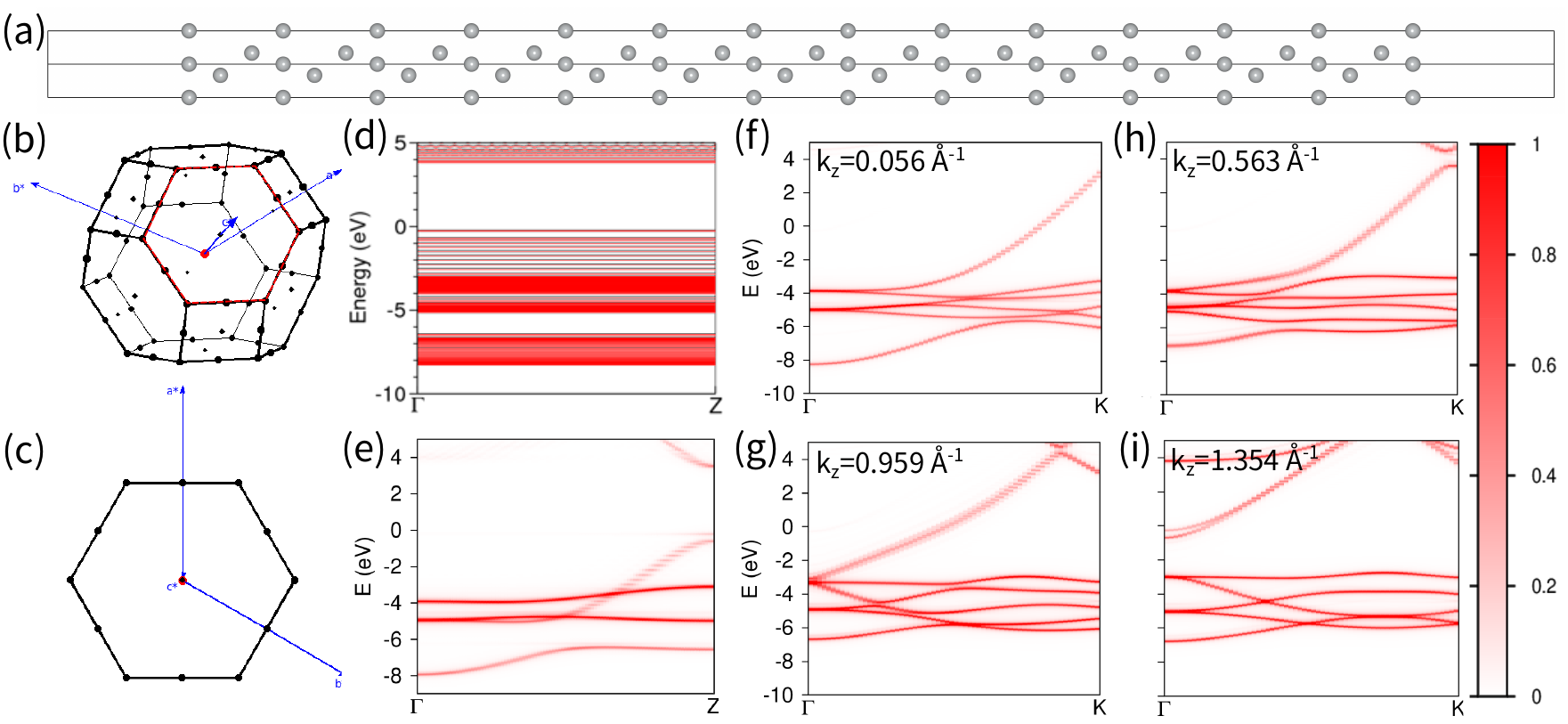}
	\caption{ $\textit{k}$-projected band structure for Ag(111) slab.(a) Geometric structure of the Ag(111) slab. (b,c) The first BZs of bulk Ag and Ag(111) slab. (d) The band structures along $\Gamma-z$. (e) The band structure by projecting the wavefunctions of the slab onto the $k$-points in the BZ of the primitive cell of Ag bulk. (f-i) The $k_z$-dependence of the band structure with bulk projection.}
  \label{FIG12}
\end{figure*}

 Figure \ref{FIG12} shows the calculation for Ag(111). We build a slab with a thickness of 113 \text{\AA}, which has 48 unit cells of Ag(111) along $z$ [Fig\ref{FIG12}(a)]. We then remove two Ag(111) layers from each side of the slab to make a surface structure. By calculating the dispersions along $\Gamma-Z$, one obtains dispersionless electronic bands [see Fig.\ref{FIG12}(d)]. By projecting the wave functions onto the $k$-points in the BZ of the primitive cell of bulk Ag [Fig.~\ref{FIG12}(e)], one obtains a band structure similar to that of the bulk. Note that the band crossing the Fermi level gets broadened as the energy goes to high energies, for which the surface effect plays a role. One can conveniently investigate the $k_z$ dependence of the band structure for the slab by projecting its wave functions onto the $k$-points in the BZ of the Ag bulk. The unfolded bands with different $k_z$s are shown in Fig.\ref{FIG12}(f-i). 
 
\subsection{Phonons of defected graphene and Moiré superlattices}
\begin{figure}
	\centering
		\includegraphics[width=1\linewidth]{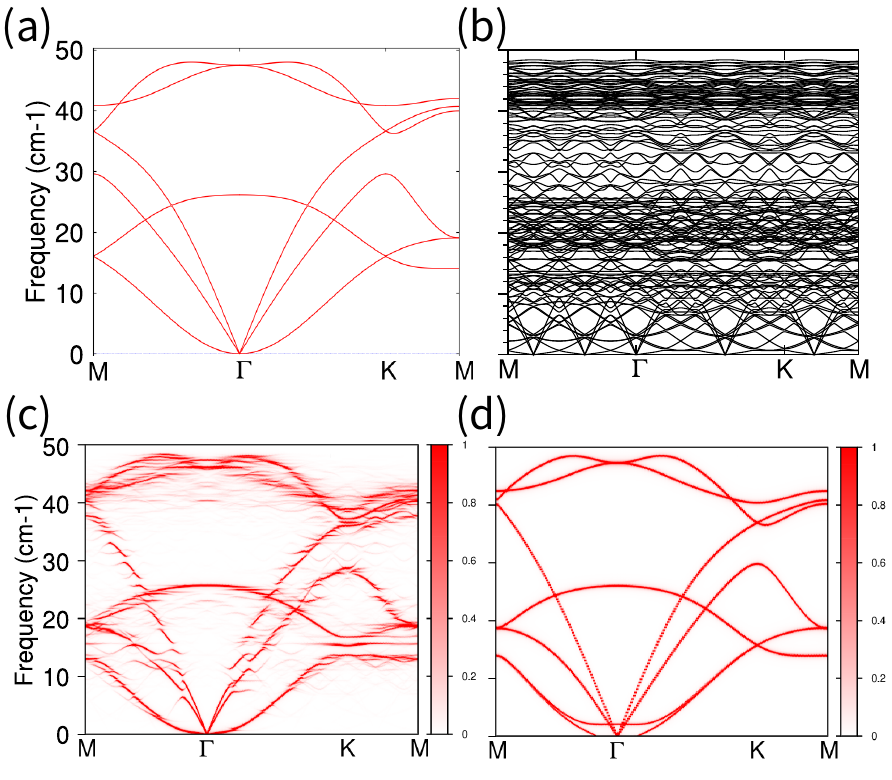}
	\caption{Unfolded phonon dispersions for the defected 5 $\times$ 5 graphene and a graphene moiré superlattice. (a) The phonon spectrum of graphene in the primitive cell. (b,c) The phonon dispersions for the defected graphene without and with band unfolding. (d) Unfolded phonon band structure with layer projection for graphene bilayer with a twist angle of 9°. }
  \label{FIG13}
\end{figure}

Our KPROJ program also allows for unfolding phonon dispersions for doped materials and interfaces. We now show its application in the defected graphene and Moiré superlattice with a twist angle of 9°. We use the small displacement method implemented in PHONOPY to obtain the phonon dispersions. The atomic forces induced by the displaced atom are calculated using LAMMPS\cite{lammps} with a machine learning potential\cite{ml1,ml2,ml3,ml4,ml5,ml6}, which was also used for the geometric relaxation. To facilitate comparison, Fig.~\ref{FIG13}(a) shows the phonon spectrum for the graphene primitive cell, and Figs.~\ref{FIG13}(b) and (c) show the folded and unfolded phonon spectrums for the defected graphene, for which the geometric structure is shown in Fig.~\ref{FIG10}(a). The unfolded phonon dispersions look similar to those for free-standing graphene. However, one can also see many minigaps that are due to the effects of carbon defects. 

We also performed layer $k$-projection for graphene Moiré superlattice, i.e., we project the eigenvectors contributed by the top layer onto the $q$-point in the BZ of the graphene primitive cell. The results are shown in Fig.~\ref{FIG13}(d). A comparison with Fig.~\ref{FIG13}(c) reveals that the twisting and stacking have a much weaker effect than do defects on the phonon dispersions of graphene.

\section{Conclusion}
In this communication, we introduce our band unfolding program KPROJ, which allows for unfolding electronic and phononic bands of materials modeled by supercells. It is interfaced with popular DFT packages and PHONONPY. The code is user-friendly and is ideal for investigating interfaces due to a highly efficient algorithm for layer projection over wave functions in a selected spatial window. Moreover, a few utilities are provided to prepare the inputs and post-processing. We have demonstrated its applications in defected graphene and heterostructures of graphene, which allows for the revealing of the effects of defect and interface on electronic and phononic dispersions. Moreover, it is useful to interface the program with the codes based on the deep learning density functional theory Hamiltonian, which allows for investigating the effects of doping and interfaces on the band structure of large systems~\cite{ml6}.

\section{Acknowledgement}
This work was supported by the National Natural Science Foundation of China (No. 12174098 to M.C.) and the US National Science Foundation (No. DMREF 2323857 to M.W.).

\bibliographystyle{elsarticle-num}
\bibliography{ref.bib}

\end{document}